\documentclass[traditabstract]{aa}
\usepackage{graphicx}
\usepackage{longtable,lscape}
\usepackage{natbib}
\usepackage{txfonts}
\usepackage{threeparttable}
\usepackage{subfigure}
\begin{document}
\newcommand{\HI}{{\ion{H}{1}}}
\newcommand{\kms}{\,km\,s$^{-1}$}
\newcommand{\cmsq}{\,cm$^{-2}$}
\newcommand{\um}{\,$\mu$m}
\def\HI{H{\,\small I}}

%
\title{A relation between circumnuclear H\,I, dust, and optical cores in low-power radio galaxies}
\titlerunning{A relation between \HI\ and dust in radio galaxies}

\subtitle{}

\author{Ilse M. van Bemmel\inst{1} \and Raffaella Morganti\inst{1,2} \and Tom Oosterloo\inst{1,2} \and Gustaaf van Moorsel\inst{3}} 
\institute{Netherlands Institute for Radio Astronomy (ASTRON), PO Box 2, 7990AA Dwingeloo, The~Netherlands,\\ \email{bemmel@astron.nl} \and Kapteyn Astronomical Institute, PO Box 800, 9700AV Groningen, The Netherlands \and National Radio Astronomy Observatory (NRAO), PO Box O, Socorro, NM 87801-0387, USA}

\authorrunning{Ilse M. van Bemmel et al.}

\date{\today}
 
\abstract{From new observations and literature data we investigate the presence of \HI, dust, and optical cores in the central kiloparsec of low-power radio galaxies. The goal of this pilot study is to identify physical relations between these components, which can help us to study kinematics and feeding mechanisms in future samples of active galaxies. Our results are consistent with neutral gas being associated with dust on sub-kiloparsec scales. Objects that have \HI\ absorption always have significant amounts of dust in their host galaxy. If there is no visible dust in the host galaxy, there is also no \HI\ absorption. The presence of an unresolved optical core correlates with the \HI\ column density, with the core being absent in high column density sources. This work opens a path for studying the kinematics of cold material in the central regions of active galaxies by combining information of \HI\ absorption and molecular lines. Consistent with previous work, we find no evidence for a compact, parsec-scale obscuring torus in low-power radio galaxies.
}
\maketitle
%

\section{Introduction} 

The central regions of active galaxies (AGN) harbour essential clues to how these amazing objects work. Differences on small scales lead to enormous variety of observed characteristics on large scales. For radio-loud AGN, a distinction is made between high- and low-power radio galaxies, referred to as FR\,II and FR\,I, respectively \citep{fr74}. Differences between these sources occur in radio morphology, optical, and X-ray properties. They are attributed to different accretion rates or accretion modes \citep[see e.g.][]{butti10}. This implies differences in kinematics and morphology of gas and dust in the central regions.

One difference between high- and low-power radio galaxies is well established. Unified schemes postulate a parsec-scale, thick dust torus to explain the bimodality in optical spectra of high-power FR\,II radio galaxies \citep[][e.g.]{pdb89,antonucci93}. In contrast, the optical nuclei of FR\,I radio galaxies appear {\em unobscured} \citep[e.g.][]{koff96,chiab99,capettiHST02}, though the host galaxies often have kiloparsec-scale dust rings and lanes \citep{dekoff00,vk02,ruiter02}. \cite{chiab99} find that the brightness of the optical core correlates with the brightness of the radio core. Therefore, they conclude that the optical core emission is non-thermal, coming from the active nucleus. This is only possible if a parsec-scale torus is absent. X-ray observations confirm this absence, and find evidence for a different accretion mode in FR\,I radio galaxies \citep{balmaverde06,hardcastle09}. 

A major question in AGN research is how the material is transported from kilo-parsec scales to the sub-parsec scale. This requires kinematical studies at high spatial and velocity resolution. Due to their proximity, FR\,I radio galaxies are best suited for this purpose. \HI\ absorption observations are a powerful tool for studying kinematics on small scales \citep[e.g.][]{vg89,morganti01,morganti03,vermeulen03,gupta06}. Molecular lines can provide essential additional information, as they also trace the cool interstellar medium. In this paper we present a pilot study to explore the relation between dust (associated with the molecular lines) and neutral gas (associated with the \HI) on kilo-parsec scales or smaller. If we can relate the gas and dust on sub-kiloparsec scales, future studies can combine information from these sources to generate a complete kinematical picture.

\begin{table*}[t]
\caption{\label{VLAobs}VLA sample description (columns 1 through 6) and observational results (columns 7 through 10).}
\begin{tabular}{lllccccccc}
\hline \hline
Name 	& ID &  $z$	& $S_{c,1.4}$	&$P_{t,4.9}$& t$_{obs}$	& $S_{abs}$	&  rms	&$\tau_{int}$& $N_{\mathrm{HI}}$	\\
B1950	&&    		& (Jy)	& (W Hz${^-1}$)	& (min)	&(mJy\,bm$^{-1}$)   & (mJy\,bm$^{-1}$) & (km s${^-1}$)	& ($\times10^{20}$\cmsq)	\\
(1)		& (2) &  (3)		& (4)		& (5)		& (6)		&   (7)	& (8)		& (9)		& (10)		\\
\hline
0055+30  & NGC315 & 	0.0165	& 0.464	& 23.84	& 27		& 27.0	& 	1.0	& 4.5$\pm$0.7	& 8.3$\pm$0.3	\\
0104+32	& 3C31 	  & 	0.0169	& 0.067	& 24.11	& 66+47	& $<$1.8& 	0.6	& $<$2.7		& $<$5.19		\\
0755+37	& 3C189	  &   0.0428	& 0.137	& 24.62	& 22		& $<$3.0	&	1.0	& $<$2.2		& $<$4.23		\\
1322+36	& NGC5141&   0.0176	& 0.054	& 23.45	& 57		& 3.3	&   0.6	& 5.5$\pm$1.0	& 10.0$\pm$1.8	\\
1346+26	& 4C26.42  &  0.0632	& 0.135	& 24.35	& 139	& 3.0	&  	0.5	& 2.2$\pm$0.4	& 4.1$\pm$0.7	\\
1350+31	& 3C293	  &  0.045 		& 1.652	& 24.93	& 20		& 148.0	& 	1.0	&12.8$\pm$0.1& 23.5$\pm$0.2\\
1626+39	& 3C338 	  &  0.0303	& 0.138	& 23.98	& 17		& $<$1.8& 	0.6	& $<$1.3		& $<$2.5		\\
2116+26	& NGC7052&  0.0156	& 0.040	& 22.80	& 88		& $<$1.2&	0.4	& $<$2.6		& $<$5.1\tablefootmark{*}		\\
\hline
\end{tabular}
\tablefoot{\tablefoottext{*}{Upper limit from our VLA observations. \citet{emonts10} set $N_{\mathrm{HI}} < 0.34 \times 10^{20}$\cmsq.}\\
Clarification of the columns: (1) B2 rado name, (2) ID used in the paper, (3) redshift, (4) core flux at 1.4\,GHz, (5) total power at 4.9\,GHz, (6) total on-source integration time, (7) peak of the absorbed flux (8) rms noise, (9) integrated optical depth, and (10) \HI\ column density.}
\end{table*}

We expand upon earlier studies by combining \HI\ information with HST imaging. Our primary goal is to determine if the presence of an optical core is associated with the absence of \HI\ absorption. Next, we search for a relation between the dust and \HI, where we expect to see absorption if dust lies in the line-of-sight to the radio source. If \HI\ is indeed associated with the dust, deeper \HI\ observations can be combined with molecular line data to study the kinematics of the central regions in a complete sample. This work is essential to provide new insight into AGN fueling and feedback.

\section{The sample} 
Our sample consists of low-power radio galaxies\footnote{Throughout this paper we use $H_{\rm 0}$= 70 km s$^{-1}$ Mpc$^{-1}$}: $log P_{\rm 1.4 GHz} < 25.5$\,W Hz$^{-1}$. They all have typical FR\,I (core-dominated) radio morphologies. The host galaxies are early-type galaxies, with a range of ellipticities and environments. For our new VLA observations we selected a representative sample of eight FR\,I radio galaxies from the HST snapshot survey of the B2 catalogue \citep{capettiHST02}. The HST snapshot images trace dust down to scales of $\sim$100\,pc and are used to identify the presence of an optical unresolved core.

To increase the sample size, we have added 14 FR\,I radio galaxies from literature. They were selected to match our criterion for radio power, and to have \HI\ data and optical HST images available. This literature sample consists of results obtained by \citet[][hereafter JvG89]{vg89}, and four additional sources from \citet{vangorkom86}, \citet{deyoung73}, \citet{jaffe94}, and \citet{hulst83}. 

\begin{figure}[t]
\centering
\includegraphics[width=80mm]{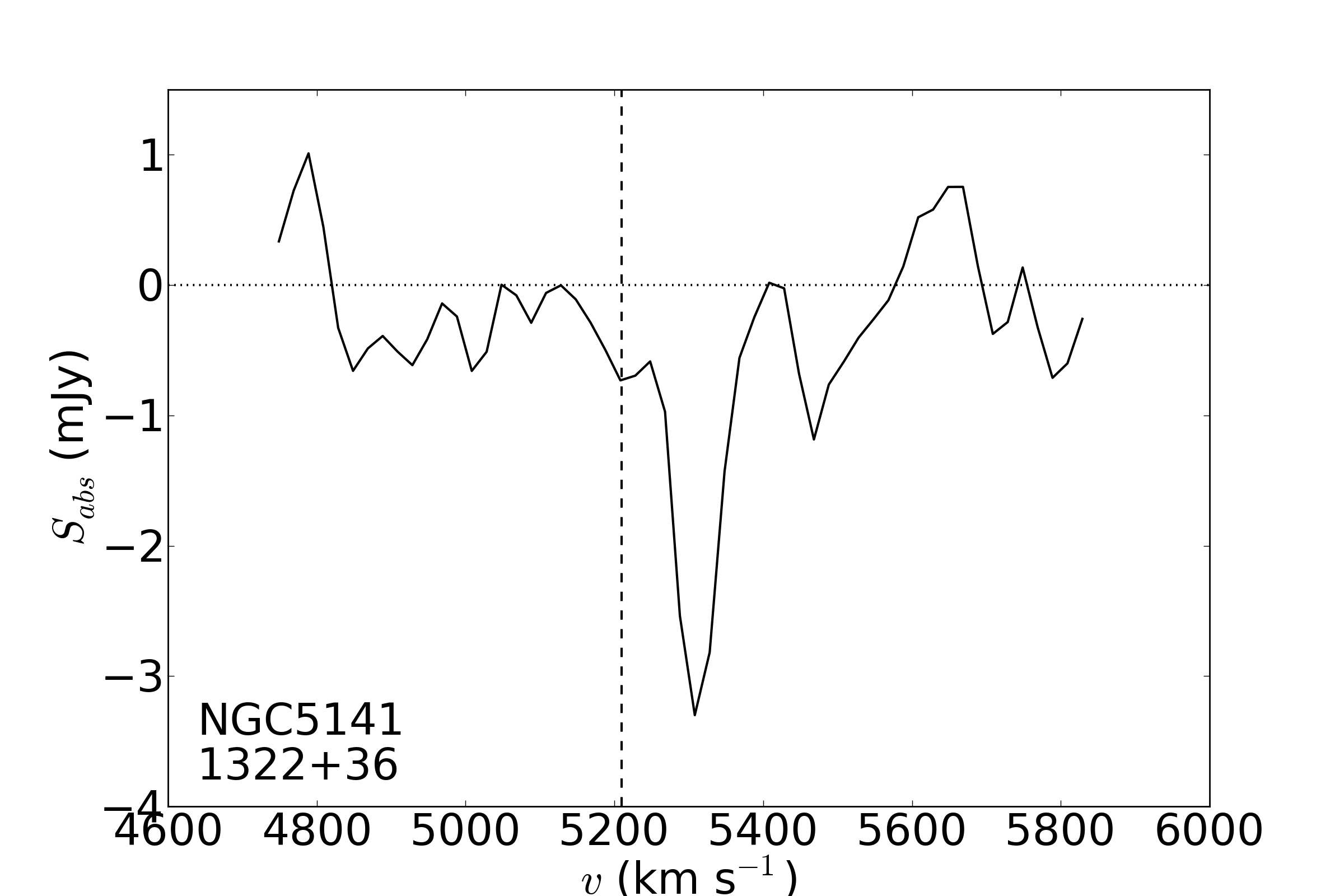}
\includegraphics[width=80mm]{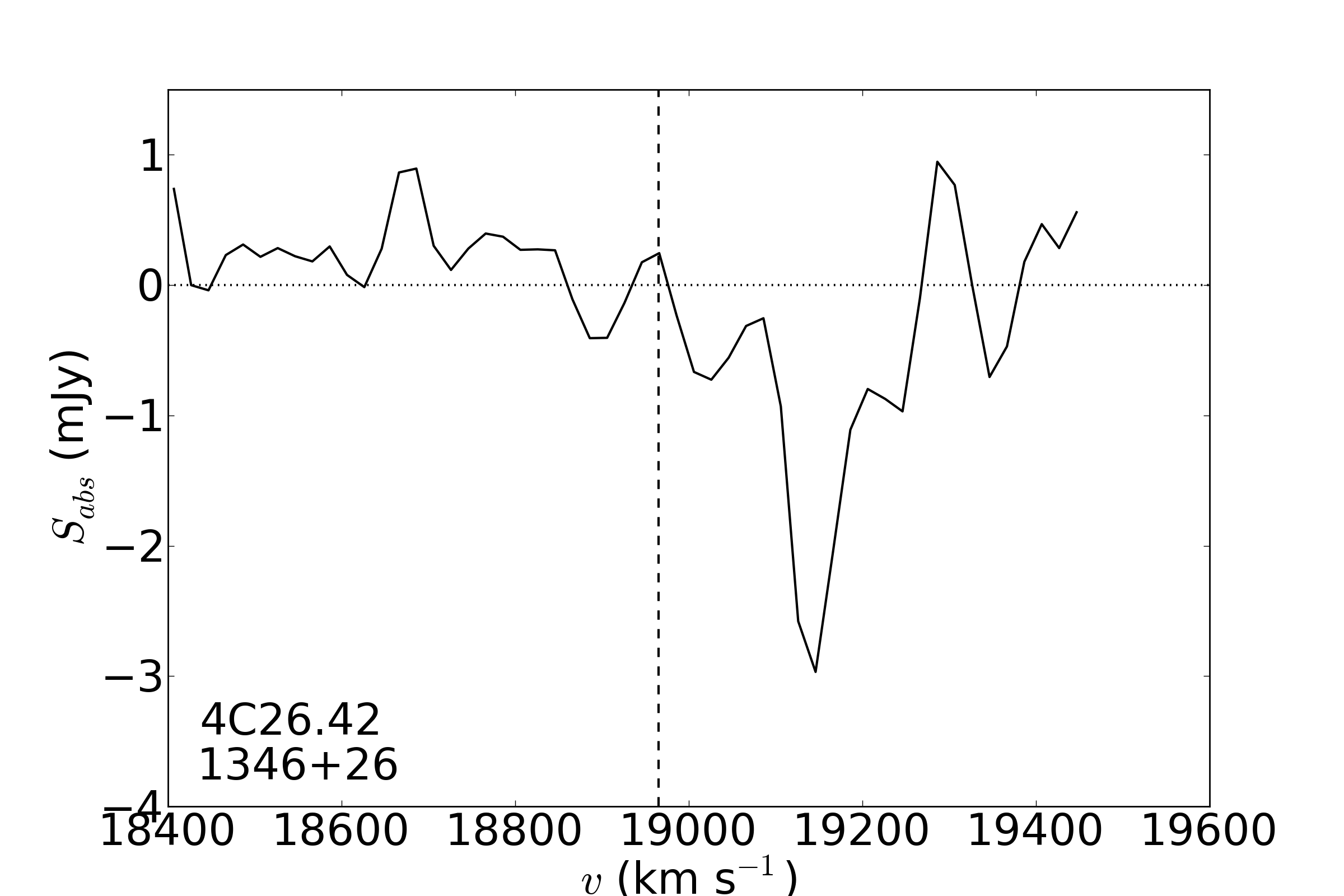}
\caption{\label{HIspec}Plot of the continuum subtracted spectra for two of the detected sources from the VLA sample. The dashed vertical line indicates the systemic velocity. The x-axis is in optical barycentric velocity. The spectra are obtained against the brightest part of the radio source, which in both of these cases corresponds to the core.}
\end{figure}

The total sample has 22 objects and is not meant to be complete, but representative of this type of radio sources. The full  sample is listed in Table~\ref{fullsample}. 

\begin{table*}[t]
\caption{\label{fullsample} \HI\ and  optical properties of the sample. First we list the sources with \HI\ absorption, then we list the non-detections. The two rows labelled as 'Total' give the total number of sources that are listed above, and the number of objects that have either dust structures or an optical core.}
\begin{tabular}{llcllcccc}
\hline \hline
\noalign{\smallskip}
B1950	& ID			& $N_{\mathrm{HI}}$ & Ref 	& Dust		& Diameter       & $i$ & $D$		& optical 	\\
 		& 			& ($\times10^{20}$\cmsq) & & morph.		& (pc) 	&$^\circ$	& (")	& core	\\
(1)		&  (2)		& (3) 	& (4)		& (5)		& (6)	& (7)		& (8)		& (9) \\
\hline
0055+30 	& NGC315	& 8.3 	&	1 & disk		& 835	& 75 & 0.15	&	y 	\\
0116+31 	& 4C31.04	& 10.8	&	2 & extended	& 2586	& 83 & 0		&	n 	\\
0238-08 	& NGC1052	&  3.0	&	3 & disk		& 416 	& 72 & 0.2	&	y 	\\
0316+41 	& 3C84		&  2.0	&	4 & extended	& 17625 	& 37 & 1.0	&	y 	\\
1146+59 	& NGC3894	& 18.2	&	2 & disk		& 911 	& 74 & 0.07	&	n 	\\
1216+06 	& 3C270		& 7.2	&	5 & disk		& 263 	& 65 & 0.07	&	y 	\\
1322-42 	& Cen A		& 50.0	&	6 & extended	& 3365	& 90 & 0		&	n 	\\
1322+36 	& NGC5141	& 10.0	&	1 & disk		& 638	& 78 & 0		&	n 	\\
1346+26 	& 4C26.42	& 4.1	&	1 & extended	& 9063	& 57 & 0.14	&	y 	\\
1350+31 	& 3C293		& 23.5	&	1 & extended	& 4130	& 67 & 0		&	n 	\\
2254-367	& IC1459		& $>$1.2	&	7 & disk		& 310	& 54 & 0		&	y	\\
\hline
Total		&			& 11 & 	&	11	&	&	& 	& 5 \\
\hline
\hline
0104+32 	& 3C31		& $<$5.2	 &	1 & disk		& 2174	& 37 & 0.8	&	y 	\\
0305+03 	& 3C78		& $<$1.47	 &	2 & none		& ...	& ... & ...	&	y 	\\
0430+05	& 3C120 		& $<$1.21	 &	2 & none	 	& ... 	& ... & ...	&	y 	\\
0755+37 	& 3C189		& $<$4.2	 &	1 & none		& ...	& ... & ...	&	y 	\\
1142+19 	& 3C264		& $<$1.65	 &	2 & disk		& 675	& 20 & 0.2	&	y 	\\
1514+07 	& 3C317		& $<$1.70	 &	2 & extended	& 2595	& $33$ & 0.25	&	y 	\\
1626+39 	& 3C338		& $<$2.5	 &	1 & extended	& 3463	& ... & 0.6	&	y 	\\
1637+82 	& NGC6251	& $<$0.3	 &	1 & disk		& 678	& 65 & 0.11	&	y 	\\
1652+39 	& 4C39.49	& $<$4.52 &	2 & none		& ...	& ...& ...	&	y 	\\
1807+69	& 3C371		& $<$2.96 &	2 & none		& ...	& ... & ...	&	y 	\\
2116+26 	& NGC7052	& $<$0.34	 &	8 & disk		& 1227	& 75 & 0		&	y 	\\
\hline
Total		&			& 11 &  	&	6	&	&		&  	&	11 \\
\hline
\hline
\end{tabular}

{\bf Notes.} Clarification of the columns: (1) B2 radio name, (2) ID used in this paper, (3) \HI\ column density (4) reference for column 3, (5) morphology of the dust, (6) size of the dust structure, (7) inclination of the disk (where applicable), (8) distance between optical core and the dust detected in the optical images, and (9) presence of an optical core: y=yes, n=no. 
\tablebib{(1) This paper;
(2) \citet{vg89};
(3) \citet{vangorkom86};
(4) \citet{deyoung73};
(5) \citet{jaffe94};
(6) \citet{hulst83};
(7) \citet{oosterloo98};
(8) \citet{emonts10}.
}
\end{table*}

\section{Observations and \HI\ results } 
subsection{VLA observations}
We observed eight sources with the VLA in A-configuration during December 2000 and January 2001. Observations were made with 6.25\,MHz bandwidth and 64 channels. The data were reduced with CASA\footnote{http://casa.nrao.edu}. We performed the default steps for \HI\ absorption line data reduction, which are described in the online manual for \HI\ absorption reduction in CASA\footnote{http://casaguides.nrao.edu}. Hanning smoothing was applied after the calibration steps. Observational parameters for each source are listed in Table~\ref{VLAobs}.

\begin{figure}[t]
\centering
\includegraphics[width=80mm]{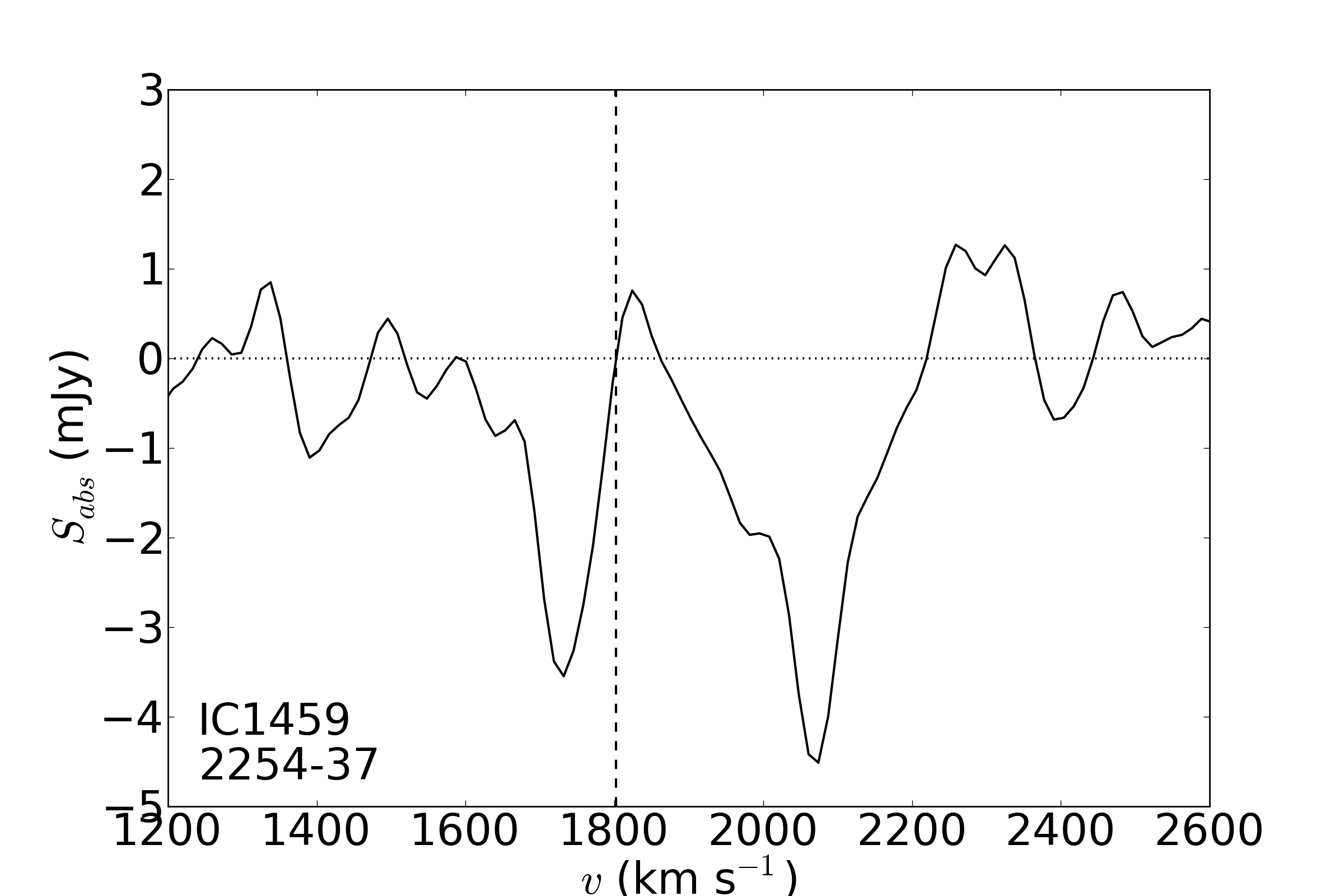}
\caption{\label{IC1459}Plot of the \HI\ absorption detected in IC\,1459 with the ATCA telescope. For comparison with the VLA spectra, the plot has been smoothed to a velocity resolution of 40\,km\,s$^{-1}$.}
\end{figure}

 \HI\ absorption is detected in four out of eight sources (see Table~\ref{VLAobs}). The peak absorbed flux exceeds $5\sigma$, with $\sigma$ being the rms noise per channel. The source 4C\,26.42 is a new detection. The $3\sigma$ absorption peak in NGC\,7052 is considered a non-detection, though it shows a hint of a resolved line. Since our observations, four sources have been studied with other telescopes. NGC\,315 is presented in \citet{morganti_n315}; 3C\,293 was observed at high resolution by \citet{beswick02}; and NGC\,5141 and NGC\,7052 were observed at lower resolution in \citet{emonts10}. Only NGC\,7052 was not detected, and a stringent upper-limit was set. Our column densities and \HI\ profiles are consistent with literature results. With our limited bandwidth and sensitivity we cannot recover the very broad component in 3C\,293 \citep{morganti03}. Two representative spectra, including 4C\,26.42, are shown in Figure~\ref{HIspec}.

\subsection{Additional data}
A detection of \HI\ absorption in IC\,1459 has been obtained with the Australia Telescope Compact Array. IC\,1459 was observed with 16.0\,MHz bandwidth, at a velocity resolution of 13\kms\ (before Hanning smoothing), and with the 375\,m configuration \citep[see Fig.~\ref{IC1459}]{oosterloo98}. The rms noise is 1.4\,mJy\,beam$^{-1}$ after Hanning smoothing. Although the spatial resolution of these observations is very poor, the background source is unresolved on arcsecond scales \citep{sadler89}. Therefore, we can assume that the absorption is coming from these small scales.

The giant radio galaxy NGC\,6251 is listed as a non-detection by JvG89. However, \citet{evans05} reported an unpublished column density of $4.5\times10^{20}$\cmsq. To confirm this, we obtained a 12 hour WSRT service observation, with 1024 channels and a bandwidth of 20\,MHz. The observation was processed similar to the VLA observations. We found no detection and set an upper limit of $N_{\mathrm{HI}}<0.3\times10^{20}$\cmsq\ for this source.

\subsection{\HI\ column density}
To derive the \HI\ column density in units of \cmsq, we used the standard equation
\begin{equation}
N_{\mathrm{HI}} = 1.835\times10^{18}  \frac{T_s}{f_c} \int{\tau({\upsilon}) d{\upsilon}} = 1.835\times10^{20}  \cdot \tau_S \cdot \Delta {\upsilon}_{\mathrm{ch}} \mbox{ ,}
\end{equation}
in which $T_S$ is the spin temperature and $f_c$ the covering fraction, $\tau_S$ is the optical depth summed over the channels where absorption is present, and $\Delta {\upsilon}_{\mathrm{ch}}$ is the velocity width of a single channel. The product $\tau_S \cdot \Delta {\upsilon}_{\mathrm{ch}}$ is the integrated optical depth, $\tau_{int}$. We assumed a spin temperature of 100\,K and a covering fraction of 1. 

For non-detections we replaced the integral with the outcome for a Gaussian profile
\begin{equation}
\int{\tau({\upsilon}) d{\upsilon}} = 1.06 \cdot \tau_p \cdot \Delta {\upsilon}_{\mathrm{FWHM}} \mbox{ ,}
\end{equation}
in which $\tau_p$ is the peak optical depth and $\Delta {\upsilon}_{\mathrm{FWHM}}$ the velocity width of the Gaussian at half maximum. In a non-detection the peak absorption depth is assumed to be less than 3$\sigma$, with $\sigma$ equal to the rms noise per channel. Since $\tau_p \approx S_{abs} / S_c$, we can replace $\tau_p$ by $3\sigma / S_c$, with $S_c$ being the core flux at 1.4\,GHz. The resulting upper limit of column density is then
\begin{equation}
N_{\mathrm{HI}} < 1.93\times10^{20} \cdot \frac{3\sigma}{S_c} \cdot \Delta {\upsilon}_{\mathrm{FWHM}} \mbox{ .}
\end{equation}

We assumed the average velocity width of the detected sources for the line width $\Delta {\upsilon}_{\mathrm{FWHM}} = 100$\,km\,s$^{-1}$. For a proper comparison, we recalculated the literature upper limits, which use different values. 

The \HI\ absorption technique is limited to objects with a bright core. Hence, in sources with a faint core, we may be missing \HI\ if it has a low column density. In our small VLA sample this bias is not obvious. In the source with the weakest core, NGC\,5141, \HI\ is detected with an integrated optical depth of $5.5\pm1.0$. The results for all VLA observations are listed in Table~\ref{VLAobs}.

Combining all the results, we have 11 detections and 11 non-detections. Owing to our selection method, these numbers cannot be used to derive detection rates for the population of low-power radio galaxies as a whole.

\begin{figure}[t]
\centering
\subfigure[Histogram of the distribution of dust.]
{\includegraphics[width=80mm]{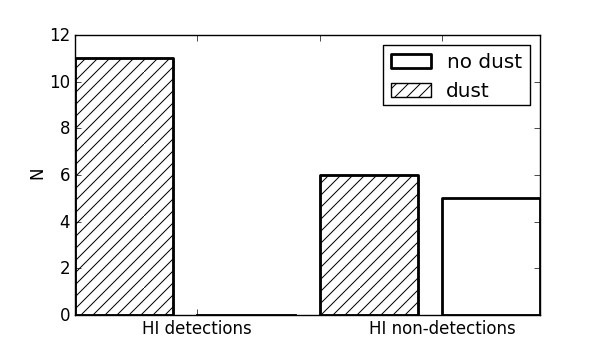}
\label{HIhista}}

\subfigure[Histogram of the distribution of optical cores.]
{\includegraphics[width=80mm]{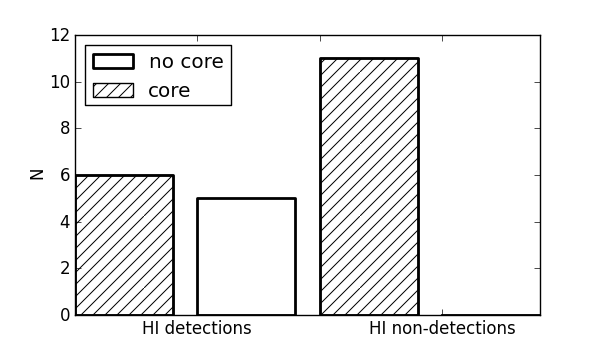}
\label{HIhistb}}

\caption{\label{HIhist}Histograms of the distributions of dust (a) and optical cores (b) in the \HI\ detections and non-detections.}
\end{figure}

\section{Results}

The goal of this study is to compare the detection of \HI\ in absorption with the presence of kiloparsec-scale dust and an optical core. 

\subsection{Relation between neutral gas, dust, and optical core}

\begin{figure}[t]
\centering
\includegraphics[width=80mm]{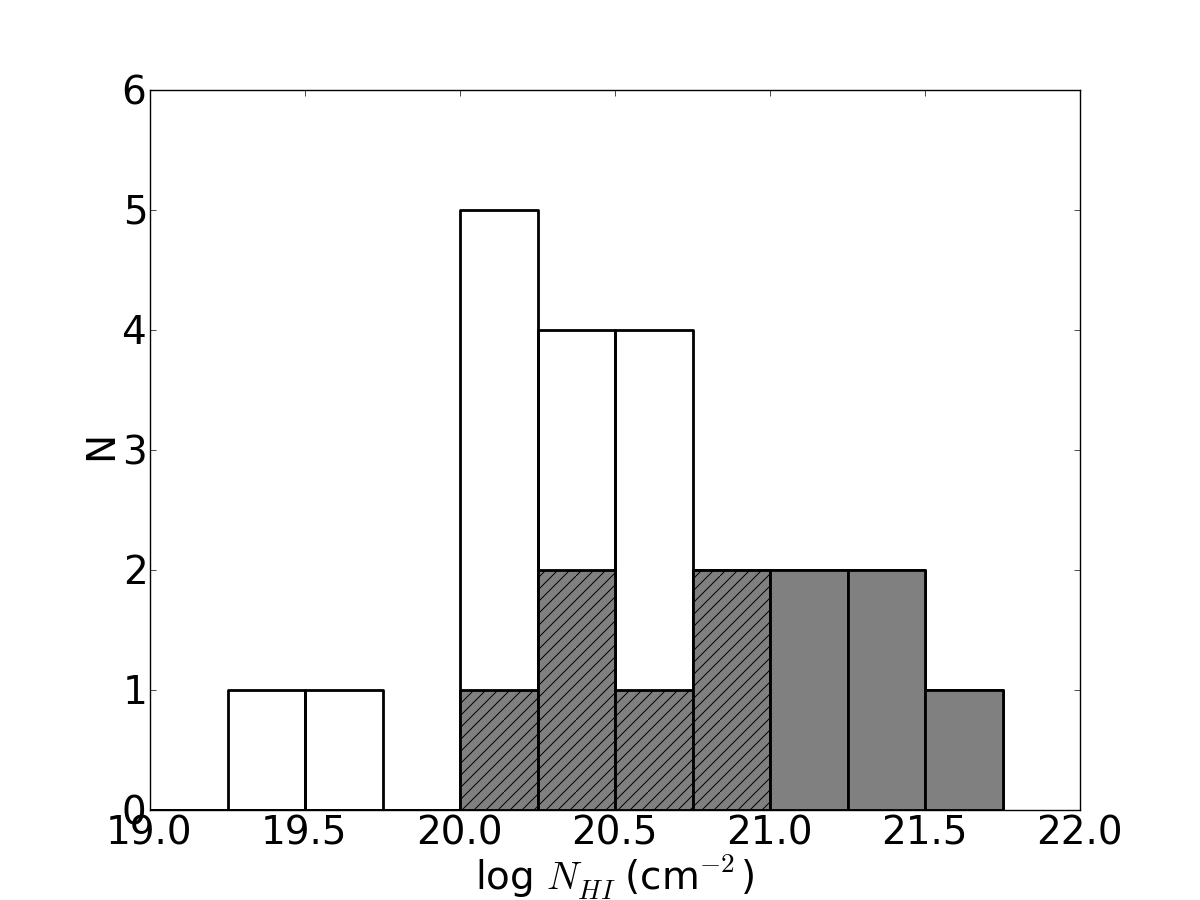}
\caption{\label{HIcolumn} Distribution of \HI\ column density for the sample. The detections are grey, the detections with an optical core are hatched. Within the detections there is a clear trend for the sources without an optical core to have a higher \HI\ column density. This is reflected in the entire sample, taking into account that the non-detections in the white area only have upper limits of \HI\ column density and all have an optical core.}
\end{figure}

The trends we find are illustrated in Figs.~\ref{HIhist} and \ref{HIcolumn} and are listed in Table~\ref{fullsample}. They can be summarized as follows:

\begin{enumerate}
\item All \HI\ absorption detections have significant amounts of dust in their host galaxies, and sources without visible dust in the host galaxies have no \HI\ absorption (Fig.~\ref{HIhista}).
\item All sources without \HI\ detection have an optical core (Fig.~\ref{HIhistb}).
\item Sources with $N_{\mathrm{HI}} > 10^{21}$\cmsq\ have no optical core (Fig.~\ref{HIcolumn}).
\end{enumerate}

We found sources with \HI\ absorption that have an optical core, see Fig.~\ref{HIcolumn}. These sources have either extended dust far from the nucleus (see section~4.2), or disks with low inclination, consistent with a line of sight through a less dense medium. If $N_H > 10^{21}$\cmsq, the unresolved optical core is not detected. This column density corresponds to a visible extinction of $\sim1$ magnitude for Galactic extinction \citep{predehl95}. Because we have no handle on the intrinsic brightness of the optical core, future work on a larger, complete sample will have to clarify the scientific merit of our third result.

\subsection{Dust morphology}
In the optical images we find three distinct morphologies for the dust: 

\begin{itemize}
\item  {\it Extended}: dust extending up to tens of kpc. In many sources this dust does not show any symmetry with respect to the core of the galaxy, often it is diffuse and/or filamentary. Some extended dust structures show evidence for smaller scale arms, or have a feather-like appearance. In a few sources it appears as a large dust lane (e.g. Centaurus~A).
\item {\it Disks}: dust in a well-defined and symmetric distribution. The dust disks are rather compact, having radii of several hundreds of parsec. We find no disks with radii larger than 1\,kpc.
\item {\it None}: in the HST images there is no visible dust. 
\end{itemize}
Our classification of dust structures is consistent with results from \citet{ruiter02}. 

The 11 sources with \HI\ absorption all have dust, and are split between extended dust (5) and disks (6). The disks are at high inclinations, with a median value of 73\degr. Extended dust either covers the optical core, or is located at less than one HST resolution element from the core. The exception is 3C\,84, where the dust is more than 1$^{\prime\prime}$ from the optical core. This source is known to have a galaxy in the line of sight. This was first discussed by \citet{minkowski57} and later confirmed by \citet{lynds70}. Two separate \HI\ absorption components are seen. The velocity differences are consistent with a component associated with the foreground galaxy and another with the radio galaxy \citep{deyoung73,rubin77}. It is not clear where the dust is located. VLBI observations are ongoing to spatially resolve the \HI\ absorption components in this complex source.

Among the 11 \HI\ non-detections, 6 sources have dust. In four cases the dust is either in a face-on disk (e.g. 3C\,31) or at large projected distance from the core. The remaining two, NGC\,6251 and NGC\,7052, have an edge-on dust disk. The non-detection of NGC\,6251 can be explained by the existence of a strong warp in the inner regions. While the outer parts of the disk are nearly edge-on at 76$^\circ$ inclination, the inner regions are at 35$^\circ$ inclination \citep{ferrarese99}. If neutral gas is present in the dust disk, this would significantly suppress the column density of \HI\ in the center. NGC\,7052 has the weakest radio core in our sample, which makes it the most difficult source to detect. Nevertheless, \citet{emonts10} set a stringent upper limit. Higher resolution observations of this source may provide insight into the structure of the disk.

In sources without dust there is no detection of \HI, and they have an unresolved optical core. Though this is consistent with the absence {\bf of} obscuration on all scales, for these five sources alone we cannot fully exclude orientation effects. Considering the sample as a whole, we would expect a tighter relation between the absence of an optical core and the presence of \HI\ absorption if a small-scale torus were present. This is not reflected in our results.

\section{Conclusions}
We have studied the presence of \HI, dust, and optical cores in the central kiloparsec of low-power radio galaxies. We find that the presence of \HI\ absorption is linked to the presence of dust. Furthermore, the {\em absence} of \HI\ absorption is linked to the presence of an optical core. The optical core is not detected if $N_{\mathrm{HI}} > 10^{21}$\cmsq. Our results are consistent with neutral gas being associated with extended dust and dust in the kiloparsec-scale disks. We require no parsec-scale dust torus.

The distribution of \HI\ may be more complicated than a regular rotating central disk. This is evident from structure in the absorption profile, and is demonstrated by several higher resolution studies. \HI\ distributed in clouds with a filling factor much lower than unity, and \HI\ in infalling and outflowing structures are seen \citep[see the case of NGC315,][]{morganti_n315}. Only VLBI observations will allow one to distinguish these effects. For sources with \HI\ absorption, the line kinematics is an essential component for understanding the feeding mechanism of active galaxies. Further insight into the physics of kiloparsec-scale disks will require deep observations of a complete sample. These are currently ongoing. More work will be done by the large surveys planned with the SKA pathfinders.

Finally, molecular line transitions can provide crucial complementary information on the kinematics of the disks. This is especially important for sources that do not have a bright, unresolved radio core. This work opens a potential path for combining molecular lines with \HI\ absorption to study kinematics on sub-kiloparsec scales. Systematic studies of the cold, molecular gas component are now starting. However, \citet{ocana10} found that radio galaxies tend to have relatively low molecular gas masses. Given the limited sensitivity of current instruments, ALMA will make a major leap in this field.

\begin{acknowledgements} 
We gratefully acknowledge discussions with Neeraj Gupta, Raymond Oonk, Preeti Kharb and Gijs Verdoes Kleijn. We thank Jacqueline van Gorkom for pointing us to relevant information on ISM physics. The VLA is operated by the National Radio Astronomy Observatory, a facility of the National Science Foundation operated under cooperative agreement by Associated Universities, Inc. This research has made use of the NASA/IPAC Extragalactic Database (NED) which is operated by the Jet Propulsion Laboratory, California Institute of Technology, under contract with the National Aeronautics and Space Administration. Based on observations made with the NASA/ESA Hubble Space Telescope, and obtained from the Hubble Legacy Archive, which is a collaboration between the Space Telescope Science Institute (STScI/NASA), the Space Telescope European Coordinating Facility (ST-ECF/ESA) and the Canadian Astronomy Data Centre (CADC/NRC/CSA). This research has made use of Aladin.
\end{acknowledgements}

\bibliographystyle{aa}

\begin{thebibliography}{33}
\expandafter\ifx\csname natexlab\endcsname\relax\def\natexlab#1{#1}\fi

\bibitem[{Antonucci(1993)}]{antonucci93}
Antonucci, R. 1993, ARA{\&}A, 31, 473

\bibitem[{Balmaverde {et~al.}(2006)Balmaverde, Capetti, \&
  Grandi}]{balmaverde06}
Balmaverde, B., Capetti, A., \& Grandi, P. 2006, A{\&}A, 451, 35

\bibitem[{Barthel(1989)}]{pdb89}
Barthel, P.~D. 1989, ApJ, 336, 606

\bibitem[{Beswick {et~al.}(2002)Beswick, Pedlar, \& Holloway}]{beswick02}
Beswick, R.~J., Pedlar, A., \& Holloway, A.~J. 2002, MNRAS, 329, 620

\bibitem[{Buttiglione {et~al.}(2010)Buttiglione, Capetti, Celotti, Axon,
  Chiaberge, Macchetto, \& Sparks}]{butti10}
Buttiglione, S., Capetti, A., Celotti, A., {et~al.} 2010, A{\&}A, 509, 6

\bibitem[{Capetti {et~al.}(2002)Capetti, Celotti, Chiaberge, de~Ruiter, Fanti,
  Morganti, \& Parma}]{capettiHST02}
Capetti, A., Celotti, A., Chiaberge, M., {et~al.} 2002, A{\&}A, 383, 104

\bibitem[{Chiaberge {et~al.}(1999)Chiaberge, Capetti, \& Celotti}]{chiab99}
Chiaberge, M., Capetti, A., \& Celotti, A. 1999, A{\&}A, 349, 77

\bibitem[{de~Koff {et~al.}(1996)de~Koff, Baum, Sparks, Biretta, Golombek,
  Macchetto, McCarthy, \& Miley}]{koff96}
de~Koff, S., Baum, S.~A., Sparks, W.~B., {et~al.} 1996, ApJS, 107, 621

\bibitem[{de~Koff {et~al.}(2000)de~Koff, Best, Baum, Sparks, R{\"o}ttgering,
  Miley, Golombek, Macchetto, \& Martel}]{dekoff00}
de~Koff, S., Best, P., Baum, S.~A., {et~al.} 2000, ApJS, 129, 33

\bibitem[{de~Ruiter {et~al.}(2002)de~Ruiter, Parma, Capetti, Fanti, \&
  Morganti}]{ruiter02}
de~Ruiter, H.~R., Parma, P., Capetti, A., Fanti, R., \& Morganti, R. 2002,
  A{\&}A, 396, 857

\bibitem[{De Young {et~al.}(1973)De Young, Roberts, \& Saslaw}]{deyoung73}
De Young, D.~S., Roberts, M.~S., \& Saslaw, W.~C. 1973, ApJ, 185, 809

\bibitem[{Emonts {et~al.}(2010)Emonts, Morganti, Struve, Oosterloo, van
  Moorsel, Tadhunter, van~der Hulst, Brogt, Holt, \& Mirabal}]{emonts10}
Emonts, B. H.~C., Morganti, R., Struve, C., {et~al.} 2010, MNRAS, 406, 987

\bibitem[{Evans {et~al.}(2005)Evans, Hardcastle, Croston, Worrall, \&
  Birkinshaw}]{evans05}
Evans, D.~A., Hardcastle, M.~J., Croston, J.~H., Worrall, D.~M., \& Birkinshaw,~M. 2005, MNRAS, 359, 363

\bibitem[{Fanaroff \& Riley(1974)}]{fr74}
Fanaroff, B.~L. \& Riley, J.~M. 1974, MNRAS, 167, 31

\bibitem[{Ferrarese \& Ford(1999)}]{ferrarese99}
Ferrarese, L. \& Ford, H.~C. 1999, ApJ, 515, 583

\bibitem[{Flaquer {et~al.}(2010)Flaquer, Leon, Combes, \& Lim}]{ocana10}
Flaquer, B.~O., Leon, S., Combes, F., \& Lim, J. 2010, A{\&}A, 518, 9

\bibitem[{Gupta {et~al.}(2006)Gupta, Salter, Saikia, Ghosh, \&
  Jeyakumar}]{gupta06}
Gupta, N., Salter, C.~J., Saikia, D.~J., Ghosh, T., \& Jeyakumar, S. 2006,
  MNRAS, 373, 972

\bibitem[{Hardcastle {et~al.}(2009)Hardcastle, Evans, \&
  Croston}]{hardcastle09}
Hardcastle, M.~J., Evans, D.~A., \& Croston, J.~H. 2009, MNRAS, 396, 1929

\bibitem[{Jaffe \& McNamara(1994)}]{jaffe94}
Jaffe, W. \& McNamara, B.~R. 1994, ApJ, 434, 110

\bibitem[{Lynds(1970)}]{lynds70}
Lynds, R. 1970, ApJ, 159, L151

\bibitem[{Minkowski(1957)}]{minkowski57}
Minkowski, R. 1957, Radio astronomy, 4, 107

\bibitem[{Morganti {et~al.}(2003)Morganti, Oosterloo, Emonts, van~der Hulst, \&
  Tadhunter}]{morganti03}
Morganti, R., Oosterloo, T.~A., Emonts, B. H.~C., van~der Hulst, J.~M., \&
  Tadhunter, C.~N. 2003, ApJ, 593, L69

\bibitem[{Morganti {et~al.}(2001)Morganti, Oosterloo, Tadhunter, van Moorsel,
  Killeen, \& Wills}]{morganti01}
Morganti, R., Oosterloo, T.~A., Tadhunter, C.~N., {et~al.} 2001, MNRAS, 323,
  331

\bibitem[{Morganti {et~al.}(2009)Morganti, Peck, Oosterloo, van Moorsel,
  Capetti, Fanti, Parma, \& de~Ruiter}]{morganti_n315}
Morganti, R., Peck, A.~B., Oosterloo, T.~A., {et~al.} 2009, A{\&}A, 505, 559

\bibitem[{Oosterloo {et~al.}(1998)Oosterloo, Morganti, \& Sadler}]{oosterloo98}
Oosterloo, T., Morganti, R., \& Sadler, E. 1999, ASPC, 163, 72

\bibitem[{Predehl \& Schmitt(1995)}]{predehl95}
Predehl, P. \& Schmitt, J. H. M.~M. 1995, A\&A, 293, 889

\bibitem[{Rubin {et~al.}(1977)Rubin, Oort, Ford, \& Peterson}]{rubin77}
Rubin, V.~C., Oort, J.~H., Ford, W.~K., \& Peterson, C.~J. 1977, ApJ, 211, 693

\bibitem[{Sadler {et~al.}(1989)Sadler, Jenkins, \& Kotanyi}]{sadler89}
Sadler, E.~M., Jenkins, C.~R., \& Kotanyi, C.~G. 1989, MNRAS, 240, 591

\bibitem[{van~der Hulst {et~al.}(1983)van~der Hulst, Golisch, \&
  Haschick}]{hulst83}
van~der Hulst, J.~M., Golisch, W.~F., \& Haschick, A.~D. 1983, ApJ, 264, L37

\bibitem[{van Gorkom {et~al.}(1989)van Gorkom, Knapp, Ekers, Laing, \&
  Polk}]{vg89}
van Gorkom, J.~H., Knapp, G.~R., Ekers, R.~D., Laing, R.~A., \& Polk, K.~S.
  1989, AJ, 97, 708

\bibitem[{van Gorkom {et~al.}(1986)van Gorkom, Knapp, Raimond, Faber, \&
  Gallagher}]{vangorkom86}
van Gorkom, J.~H., Knapp, G.~R., Raimond, E., Faber, S.~M., \& Gallagher, J.~S.
  1986, AJ, 91, 791

\bibitem[{Verdoes-Kleijn {et~al.}(2002)Verdoes-Kleijn, Baum, de~Zeeuw, \&
  O'Dea}]{vk02}
Verdoes-Kleijn, G., Baum, S.~A., de~Zeeuw, P.~T., \& O'Dea, C.~P. 2002, AJ,
  123, 1334

\bibitem[{Vermeulen {et~al.}(2003)Vermeulen, Pihlstr{\"o}m, Tschager, de~Vries,
  Conway, Barthel, Baum, Braun, Bremer, Miley, O'Dea, R{\"o}ttgering,
  Schilizzi, Snellen, \& Taylor}]{vermeulen03}
Vermeulen, R.~C., Pihlstr{\"o}m, Y.~M., Tschager, W., {et~al.} 2003, A{\&}A,
  404, 861

\end{thebibliography}

\end{document}